\documentclass[conference]{IEEEtran}
\IEEEoverridecommandlockouts
\usepackage{cite}
\usepackage{amsmath,amssymb,amsfonts}
\usepackage{algorithmic}
\usepackage{graphicx}
\usepackage{textcomp}
\usepackage{xcolor}
\usepackage{url}
\usepackage{enumitem}
\def\BibTeX{{\rm B\kern-.05em{\sc i\kern-.025em b}\kern-.08em
    T\kern-.1667em\lower.7ex\hbox{E}\kern-.125emX}}
\begin{document}

\title{\textbf{Audio-to-Score Conversion Model \\ Based on Whisper methodology}}

\author{
\IEEEauthorblockN{
\textit{Hongyao Zhang\textsuperscript{1}}, \textit{Bohang Sun\textsuperscript{2}}}
\\
\IEEEauthorblockA{\textsuperscript{1}College of Information Science and Engineering, East China University of Science and Technology, Shanghai, China}
\IEEEauthorblockA{\textsuperscript{2}School of Information and Software Engineering, University of Electronic Science and Technology of China, Chengdu, China}

}

\maketitle

\begin{abstract}
This thesis develops a Transformer model based on Whisper, which extracts melodies and chords from music audio and records them into ABC notation. A comprehensive data processing workflow is customized for ABC notation, including data cleansing, formatting, and conversion, and a mutation mechanism is implemented to increase the diversity and quality of training data. This thesis innovatively introduces the ``Orpheus' Score", a custom notation system that converts music information into tokens, designs a custom vocabulary library, and trains a corresponding custom tokenizer. Experiments show that compared to traditional algorithms, the model has significantly improved accuracy and performance. While providing a convenient audio-to-score tool for music enthusiasts, this work also provides new ideas and tools for research in music information processing.
\end{abstract}

\begin{IEEEkeywords}
audio-to-score, abc notation, whisper model, transformer model
\end{IEEEkeywords}

\section{INTRODUCTION}
With the burgeoning development of the digital music industry, a vast amount of music data in the form of audio files has emerged. Despite the increasing complexity in music classification, retrieval, recommendation, and content analysis technologies, a significant barrier remains for music educators and amateur music enthusiasts due to the lack of available music sheets for a vast number of songs. This highlights the gap in music transcription technology. While professionals can recognize melodies and chords easily through practice and experience, for music enthusiasts and beginners, the process of transcribing music is costly and inefficient. Therefore, the development of an automated audio-to-sheet music tool holds great significance for both music enthusiasts and the field of music education.

Currently, in the field of audio processing, digital signal processing and waveform analysis are commonly used to extract musical information\cite{an_overview}. With the advancement of machine learning and deep learning technologies, new algorithms have also been applied to the field of music information processing, such as neural network-based music generation and music classification\cite{deep_learning}. However, these methods have the following drawbacks: 1) Difficulty in obtaining data, with high costs associated with annotating music data\cite{deep1}; 2) Poor generalization capabilities of the models, making it challenging to adapt to different musical styles\cite{challenges}; 3) Algorithms generally only extract rhythm\cite{beat_transformer} or the main melody or chords only\cite{chord_transformer}, and it is difficult to extract both melody and chords simultaneously.

The ``Orpheus' Score" notation method proposed in this thesis can encompass both melody and chord information, and, in conjunction with the Whisper pre-trained model, can deduce the music sheet.

The research presented in this thesis is of significant importance to the analysis and processing of digital music. It can assist in composition creation, pitch analysis and correction, music retrieval, and music sentiment analysis. Additionally, this model can also provide high-quality data sources for other research endeavors.

\section{DATASET BUILDING}
In the task of music score extraction, both tone and beat are important. Previous studies have not dealt with beat well in deep learning. Since the Whisper model recently showed success in weakly supervised learning\cite{whisper}, we intend to try to prepare datasets and train the model through this way.
Just like the whisper model, using 680k hours of data for weakly supervised learning, we get some music and ABC scores from Internet. In order to ensure that the keys of different scores are consistent, we ensure that the time values represented by notes are unified by converting the time values of beats. In order to expand the data, we disassemble each section and re-splice it. Then, in order to further expand the data, we replace some pitches, and extend some beats, from the perspective of ``avoid affecting harmony".

To successfully construct and train the models, it is pivotal to convert raw audio data into a suitable input format. The workflow includes data cleansing, data formatting, mutation, and data conversion, with the aim of supplying superior quality data inputs for subsequent model training. Also, we design a unique notation called ``Orpheus' Score" for modeling tokenization. Through case studies and algorithm validation, we substantiate the effectiveness of this workflow in generating music.

\subsection{Data Source}
The main data source utilized in this research is the extensively used ABC music notation format, esteemed by musicians and researchers due to its simplicity and versatility. The dataset comprises a varied range of scores, embracing various musical styles and complexities, to ensure the diversity required for model training. Initially, we got around 3000 ABC notations of pop songs from website. 

\subsection{Data Cleansing}
Data cleansing is the initial phase in the entire processing workflow. By employing regular expressions, we precisely identify and eliminate irrelevant information and noise, primarily from the score's metadata. The precise cleansing methods undertaken include:
\subsubsection{Removing Metadata Rows}This includes excluding lines containing details like the title (T:), composer (C:), and meter (M:) to yield clean note information. To make training easier, we turn all the notes into C-Key, set Meter into 4/4, and set the meter of one note to 1/4, so that all the notes are formatted.
\subsubsection{Eliminating Invalid Markers}For lines that contain meter marks and additional symbols, data normalization is executed through regular expressions, refining the score format.
\subsubsection{Handling Measures}Due to the error of processing and the error of the original data itself, the total time value of some sections may be inconsistent with the time sign. Therefore, these sections need to be checked and corrected. For the syllables whose total duration of notes is less than the supposed length of a bar, it is necessary to add an empty beat; for those whose total time value of notes exceeds the supposed length of bars, they are directly discarded.

\begin{figure}[htbp]
    \centering
    \includegraphics[width=0.5\linewidth]{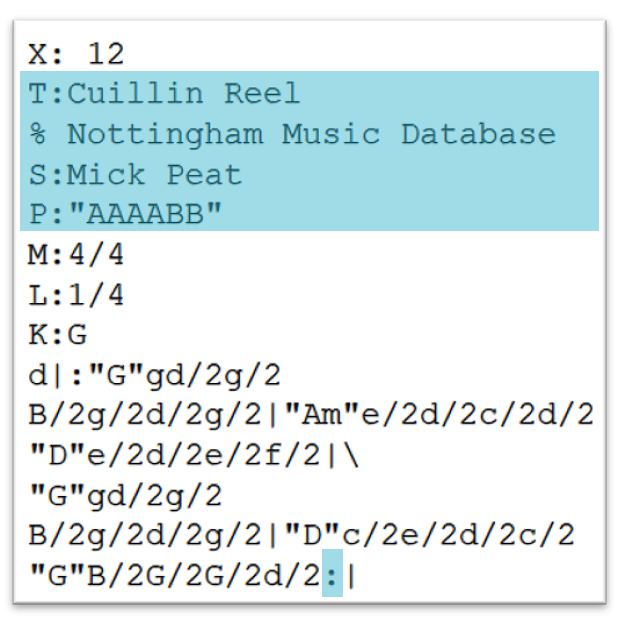}
    \caption{Data Cleansing}
    \label{fig:data_cleansing}
\end{figure}

\subsection{Data Formatting}
Due to the inconsistency of key signatures (K), minimal note duration (L), and rhythmic divisions (M) in each musical score, it becomes essential for facilitating the training of our model to unify these scores in a uniform format. The key signature (K) will be universally transposed to C major, the shortest note duration unit (L) will all conform to 1/192(considering triple notes), and the rhythmic divisions (M) will all transform into 192/192. At last, each syllable would contain notes and chord root tones with the corresponding tonal pitch in C major key, and the duration after each pitch would be converted into the equivalent duration in 4/4 beats.

\begin{figure}[htbp]
    \centering
    \includegraphics[width=0.75\linewidth]{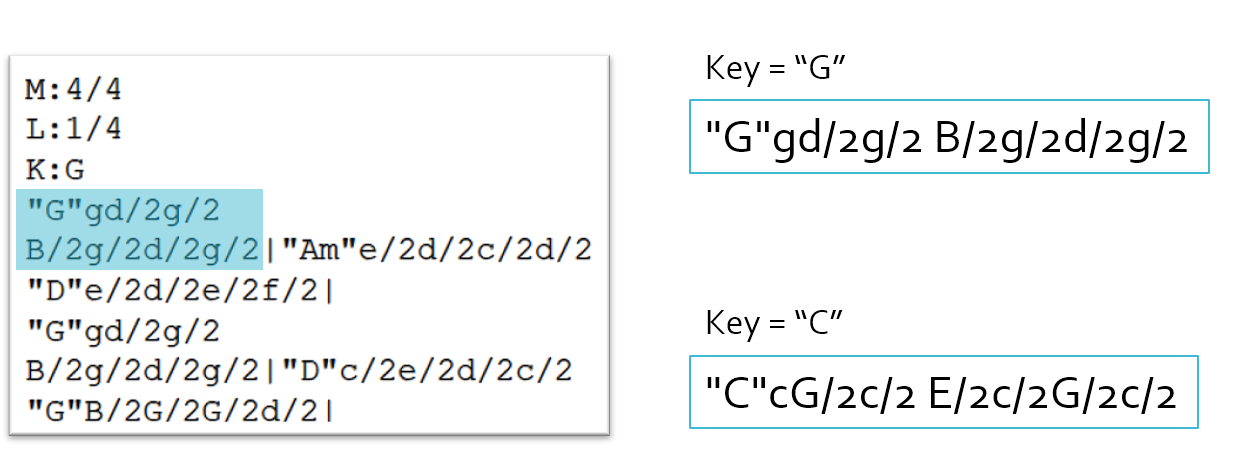}
    \caption{format the key}
    \label{fig:data_format_key}
\end{figure}

\begin{figure}[htbp]
    \centering
    \includegraphics[width=0.75\linewidth]{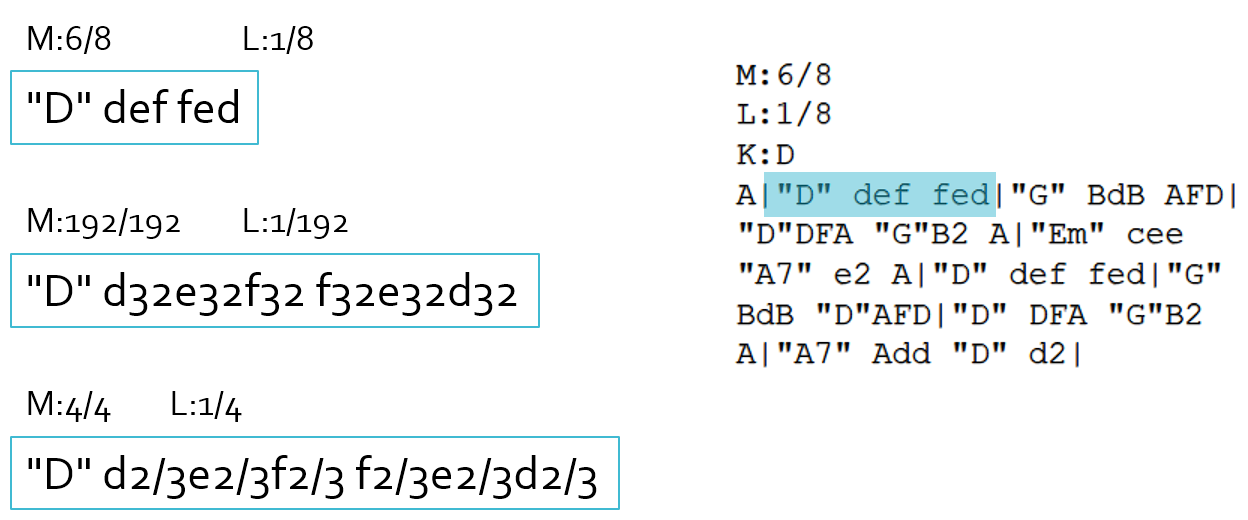}
    \caption{format the meter}
    \label{fig:data_format_meter}
\end{figure}

\subsection{Mutation Mechanism}
We introduce a mutation in our dataset. The goal is for the model to learn single notes rather than entire measures. Introducing a mutation mechanism allows for the generated notes to experience random shifts in pitch, thereby amplifying the diversity and originality of musical compositions. Through the generation and scrutiny of random numbers, the program is capable of simulating the unpredictability intrinsic in musical creation, aiding to generate more expressive musical segments. Furthermore, the conditional checks and logging incorporated into the code ascertain that the notes remain within a valid range and facilitate subsequent analysis. 

\begin{figure}[htbp]
    \centering
    \includegraphics[width=0.75\linewidth]{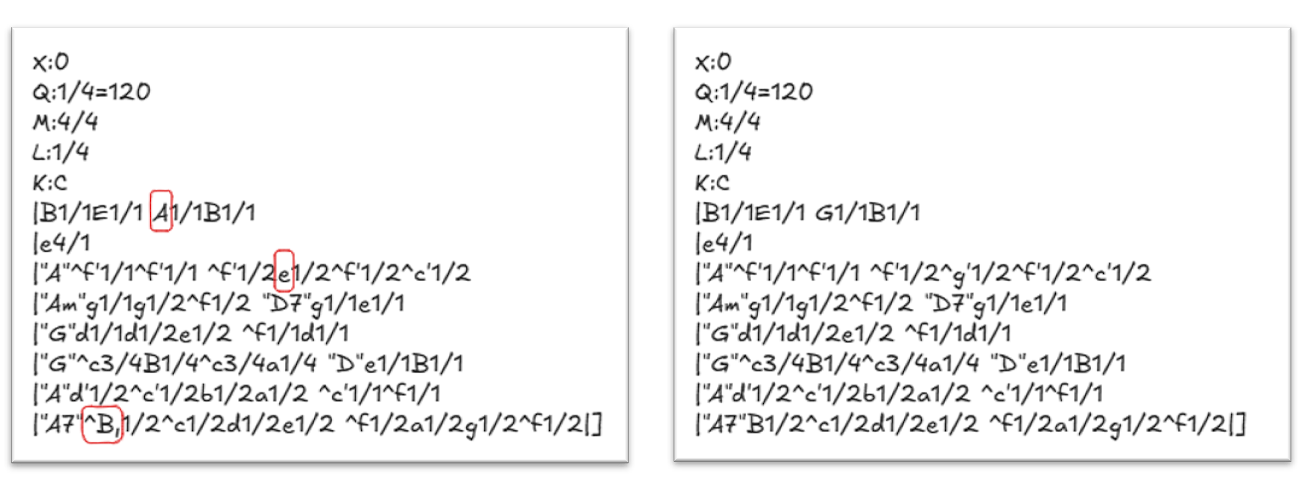}
    \caption{mutation}
    \label{fig:mutation}
\end{figure}

Since the so-called ``harmonious" music comes from the real distribution of the world, our replacement involves Gaussian sampling, which replaces part of the notes in a reasonable proportion from a statistical point of view, so as to enhance the diversity of data and expand the collective volume and quality of data.

\subsection{Database of Musical Scores Generated by Randomly Combined Sections}

To train the model with sufficient ABC scores, we compiled a database by randomly combining previously created sections. There are two methods to chose: random sampling, where each section had an equal chance of selection; or Gaussian sampling, which we adopted due to the Gaussian distribution observed in real musical scores, particularly in the frequency of notes and chords. By employing Gaussian sampling, we aimed to better align the matrix space of our training data with the nuances of human-created musical scores. Our dataset comprises 150,000 scores, each generated from an 8-section combination selected via Gaussian sampling.

By comparison, we found that the model performance improved by 13.1 percent. Thus we chose to apply the Gaussian Sampling.
This method ensures the diversity and balance of the data.

\subsection{Data Conversion}
In the development of an audio-to-score model based on the Transformer architecture, particularly for the conversion of audio to ABC notation, a specialized Tokenizer is essential. Standard natural language processing Tokenizers are ill-suited for handling the unique symbols and structures of musical notation. Our ``Orpheus' Score" addresses this by meticulously assembling a comprehensive symbol library, covering all elements from basic notes (e.g., C, D\#, Eb) to complex chord markings (e.g., C7, Am) and rhythmic patterns (e.g., 1/4, 1/8).

\begin{figure}[htbp]
    \centering
    \includegraphics[width=0.75\linewidth]{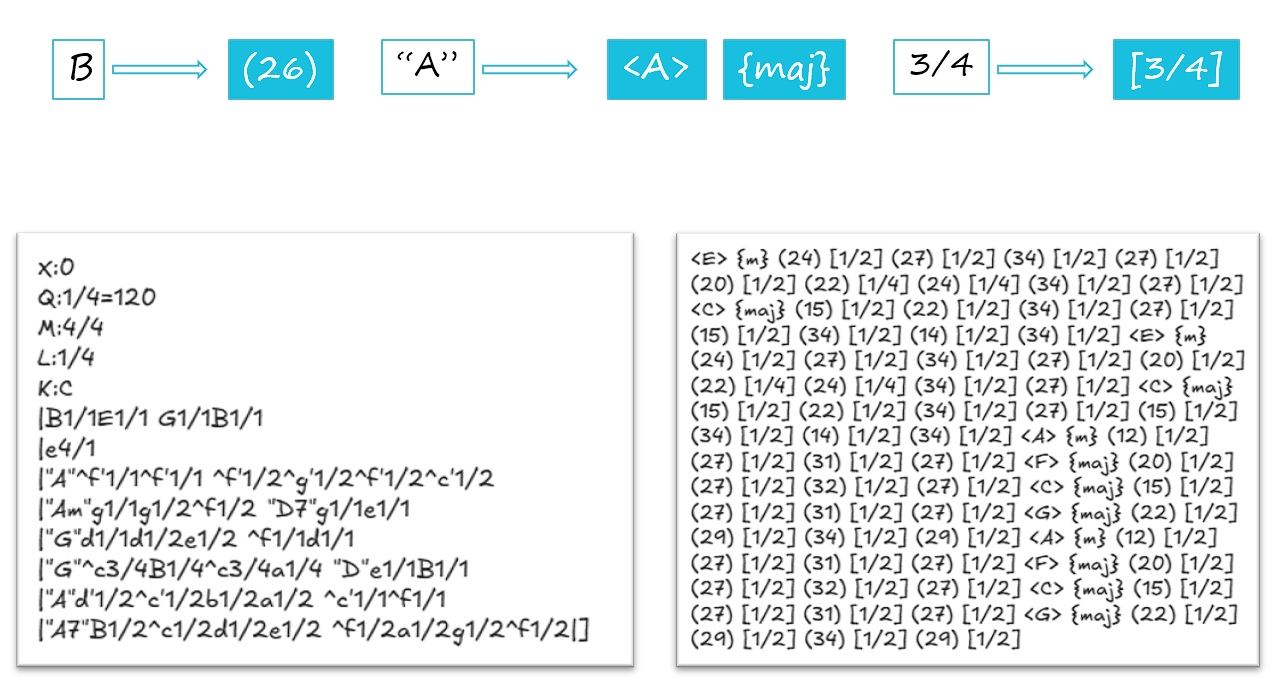}
    \caption{Tokenize into Orpheus' Score}
    \label{fig:tokenize}
\end{figure}

Each element in the musical language is assigned a unique encoding, creating a mapping table that allows the Transformer model to understand and generate the musical vocabulary. This significantly enhances the model's ability to comprehend and produce musical expressions. During the music score extraction process, ABC notation scores are converted into encoding sequences defined by "Orpheus' Score". These sequences, along with the original songs, are used to train the Transformer model. In the generation phase, the model outputs encoded sequences, which are then mapped back to ABC notation symbols to produce the transcribed score of the original song.

\subsection{Training set generation}
By processing the above, a training dataset based on the correspondence between ABC notation and MIDI/WAV formats is established. Initially, ABC notation is converted to MIDI, followed by the conversion of MIDI to WAV, yielding a WAV file. This dataset comprises features in the ``Orpheus' Score" format and labels for WAV files, facilitating subsequent model training to enhance performance and accuracy. Considering training cost and efficiency, the dataset comprises 150,000 data points, each containing 8 sections. The audio files are sampled at 16,000 Hz and last approximately 9 seconds.

\section{MODEL TRAINING}
\subsection{Model Selection}
From an input-output standpoint, the model presented in this thesis takes an audio sequence as input and produces the corresponding ``Orpheus' Score" sequence, thus constituting a ``sequence-to-sequence" model. The Transformer model is well-suited for this task due to its robust sequence processing capabilities\cite{transformer1}. Consequently, the Transformer model is the preferred choice for training in this thesis.

The Whisper model, another Transformer-based model, has demonstrated strong performance in speech recognition. Given the analogous input-output structures between speech recognition tasks and music generation tasks (both involving different tones), this thesis selects the Whisper pre-trained model as the foundation. By customizing its tokenizer training, it is adapted for the music recognition task.

\subsection{Model Reconstruction}
Given that Whisper's training is centered on speech-to-subtitle conversion, there may be some synergy with the music-to-score task, but this thesis concentrates on the more specialized domain of music-to-score conversion. Additionally, Whisper's support for multiple languages results in a substantial vocabulary, potentially impacting the model's training process due to the large output layer. Furthermore, the tokenizer in Whisper lacks specific information pertinent to music notation, necessitating a redefinition of the mapping between tokens and musical symbols. To address these challenges, this thesis opted to discard the existing model and pre-trained weights, opting instead to construct a model and tokenizer tailored for music notation from the ground up for comparative analysis. Subsequent sections of this thesis will elaborate on this approach.

\subsection{Model design}
In our training process, we utilized the Whisper-provided Whisper Feature Extractor to first convert audio inputs into mel-spectrograms. These spectrograms were then downsampled to approximately 3000 tokens. We employed Cross-Entropy Loss for training and used FP16 mixed precision to optimize performance. The warmup ratio was set to 0.1, and we saved the model weights with the lowest Word Error Rate (WER) at each epoch.
The WER is defined as:
\[
\text{WER} = \frac{S + D + I}{N}
\]
where $S$ indicates the number of substitutions, $D$ indicates the number of deletions,  $I$ is the number of insertions, and $N$ indicates the total number of words in the reference. 

Due to the inherent complexity of music transcription task, such as the relatively low information density in music and the stringent requirements for accurate musical notation, it is common for the WER in music transcription tasks to fall within the range of 30\% to 50\%.

Additionally, we utilized a customized vocabulary for tokenization with word-level processing. This necessitated a substantial output layer, leading us to modify the model head from $A \in \mathbb{R}^{512 \times 51863}$ to $A \in \mathbb{R}^{512 \times 135}$. This adjustment significantly reduced both training difficulty and parameter count, facilitating the exploration of various model configurations and enabling deployment on edge devices.

Initially, we trained a small model comprising 2 encoders and 2 decoders, designed to be operable on edge devices such as mobile phones or Raspberry Pi. Following this, we trained a more advanced model with 4 encoders and 4 decoders, which demonstrated the best performance among the configurations tested. In our third iteration, we scaled up the model further by increasing the number of encoders and decoders to 6 each. This larger model has a parameter count in the range of several hundred thousand.

We evaluated several architectures, which exhibited similar performance with minor differences, as detailed in the accompanying tables. Among the models tested, we recommend the encoder * 4 + decoder * 4 architecture, trained from scratch. While it did not achieve the lowest eval loss or WER, it provided the most favorable perceptual quality.

In our experiments, we trained five different model configurations from scratch: \par(1) a model with 2 encoders and 2 decoders (where we adjusted the feedforward layer dimension to 768 from the original 1536)

\par(2) a model with 4 encoders and 4 decoders
\par(3) a model with 6 encoders and 6 decoders
\par(4) a model with a Whisper\_small encoder (initialized with pre-trained weights) and 4 decoders
\par(5) a model with a Whisper\_small encoder (initialized with pre-trained weights and kept fixed during training) and 4 decoders.
\par

\subsection{Evaluation}
All five models achieved satisfactory results, though their convergence rates varied. Notably, the model with 4 encoders and 4 decoders consistently delivered the best auditory quality, leading us to recommend this configuration for optimal performance. Additionally, while the Whisper\_small encoder (fixed) + 4 decoders model did not show substantial improvements in performance, it demonstrated faster convergence during training and maintained accuracy without significant decline. This suggests that the encoding capabilities of the Whisper-trained encoder may generalize to musical data to some extent, aligning with our expectations. Detailed performance metrics are provided in the tables and figures below.

\begin{table}[h]
\caption{Comparison of Model Architectures}
\centering
\label{tab:model_comparison}
\begin{tabular}{|ll|r|r|r|}
\hline
\multicolumn{2}{|l|}{\textbf{Model Architecture}} & \textbf{\#Paras} & \textbf{Loss/wer} & \textbf{Infer Speed} \\
Encoder & Decoder & & & (CPU/CUDA) \\
\hline
$\times$ 2 & $\times$ 2 & 9.4M & 0.01/47.2 & 0.37 s/0.23 s \\
\hline
$\times$ 4 &  $\times$ 4 & 17.9M & 3.66e-3/46.2 & 0.23 s/0.16 s \\
\hline
$\times$ 6 & $\times$ 6 & 26.2M & 3.21e-3/48.07 & 0.38 s/0.23 s \\
\hline
$WS^{*}$  & $\times$ 4 & 105.9M & 4.31e-3/47.0 & 0.45 s/0.18 s \\
\hline
$WS^{*}$ (fixed) & $\times$ 4 & 105.9M & 5.54e-3/46.8 & 0.45 s/0.18 s \\
\hline
\multicolumn{2}{|l|}{$WS^{*}$ (Orig)} & 244M & - & -\\ 
\hline
\multicolumn{5}{l}{ $^{*}$ ``WS" represents Whisper\_small model }
\end{tabular}
\end{table}

We conducted speed tests on both CUDA and CPU platforms. Initially, a warm-up phase with empty data was performed to ensure full model loading. Subsequently, we executed inference 100 times on the same audio data and computed the average inference speed for each model. Given that the CPU used was a Core i9, the performance differences may be less pronounced. Moreover, the computational overhead of Transformers is influenced not only by the number of parameters but also significantly by the number of tokens, due to the size of the ${Q, K, V}$ matrices.

\begin{figure}[!htbp]
    \centering
    \includegraphics[width=0.5\linewidth]{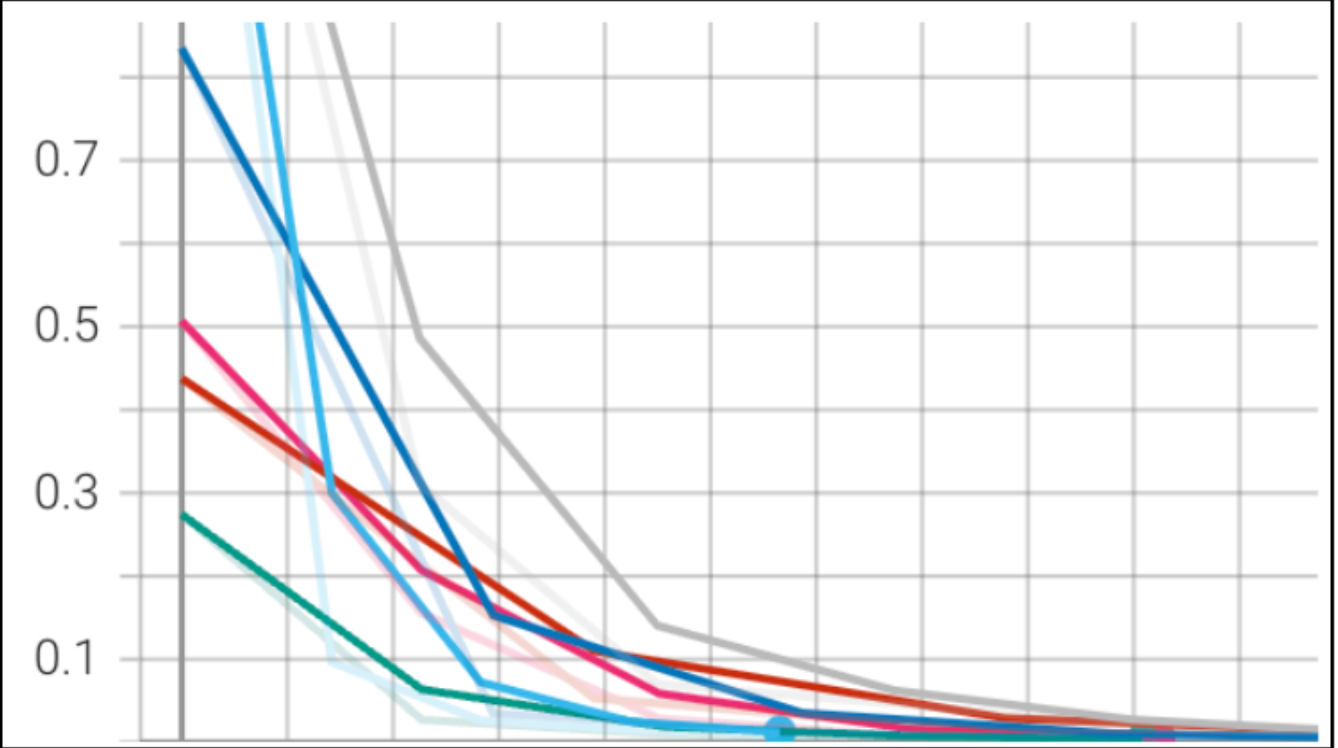}
    \caption{eval/loss}
    \label{fig:loss_figure}
\end{figure}
\begin{figure}[!htbp]
    \centering
    \includegraphics[width=0.5\linewidth]{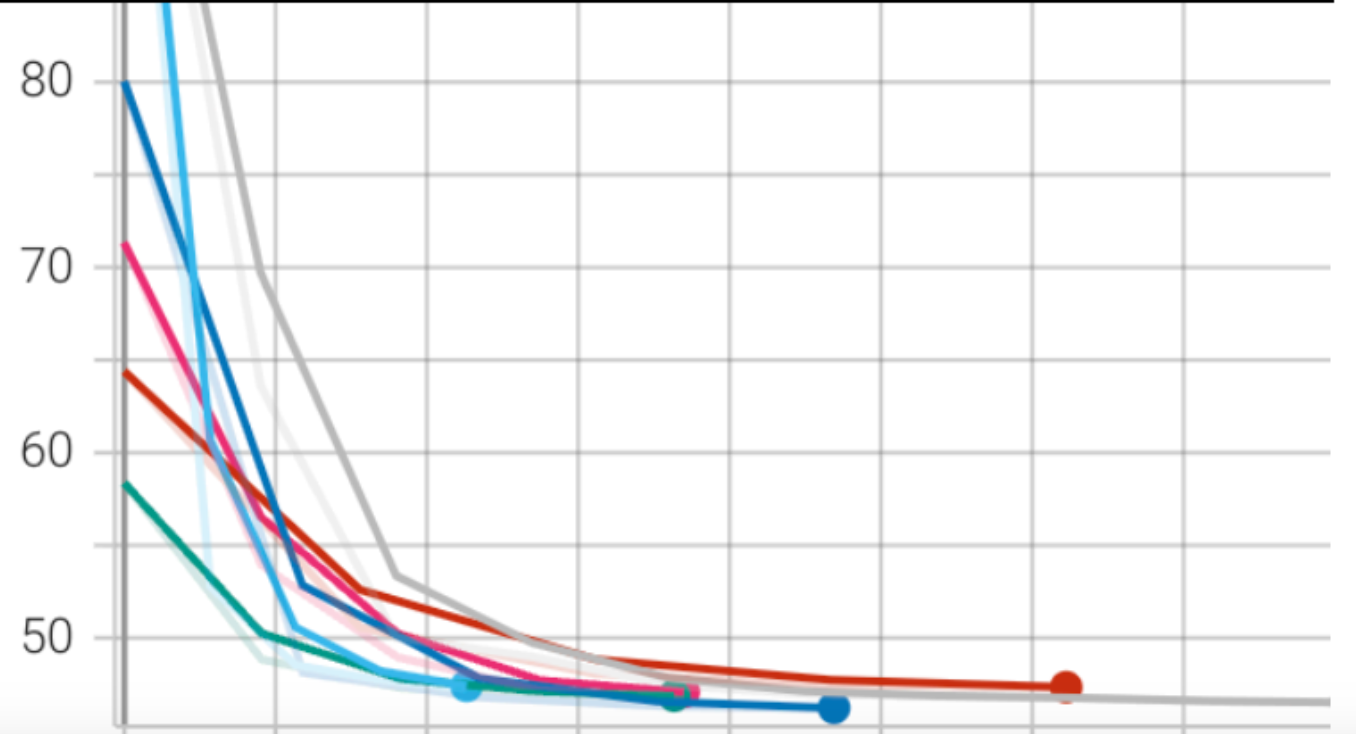}
    \caption{eval/wer}
    \label{fig:wer_figure}
\end{figure}

\section{CONCLUSIONS}

In summary, this thesis has successfully developed a novel model based on the Whisper architecture, which accurately converts audio files into ABC notation. By carefully proposing and evaluating five distinct methods, we have demonstrated the effectiveness and versatility of our approach. The creation of a specialized database has significantly enhanced the model's capabilities, guaranteeing precise and trustworthy conversions. This research not only provides music enthusiasts with an easy-to-use audio-to-score conversion tool but also contributes new insights and resources to the field of music information processing. Additionally, we have made our data openly accessible\footnote{Open source dataset address: \\ \url{https://huggingface.co/datasets/BOB12311/Orpheus_Hearing}}, aiding related research efforts and fostering collaborative advancement in the field.

\newpage
\bibliographystyle{unsrt}
\bibliography{bib-template}
\vspace{12pt}

\end{document}